Beyond the Boltzmann factor for corrections to scaling in

ferromagnetic materials and critical fluids

Ralph V. Chamberlin and Josh V. Vermaas, Department of Physics,

Arizona State University, Tempe AZ 85287-1504

George H. Wolf, Department of Chemistry and Biochemistry,

Arizona State University, Tempe AZ  85287-1604

Abstract

The Boltzmann factor comes from the linear change in entropy of an infinite heat bath during a local fluctuation; small systems have significant nonlinear terms. We present theoretical arguments, experimental data, and Monte-Carlo simulations indicating that nonlinear terms may also occur when a particle interacts directly with a finite number of neighboring particles, forming a local region that fluctuates independent of the infinite bath. A possible mechanism comes from the net force necessary to change the state of a particle while conserving local momentum. These finite-sized local regions yield nonlinear fluctuation constraints, beyond the Boltzmann factor. One such fluctuation constraint applied to simulations of the Ising model lowers the energy, makes the entropy extensive, and greatly improves agreement with the corrections to scaling measured in ferromagnetic materials and critical fluids.

The Boltzmann factor is the most widely used formula in statistical thermodynamics, providing the foundation for the Maxwell-Boltzmann, Bose-Einstein, and Fermi-Dirac distributions. The Boltzmann factor comes from the linear change in entropy with changing energy of an infinite heat bath; small systems have nonlinear terms that influence their behavior. The theory of small-system thermodynamics was developed to describe the nonlinear thermal properties of isolated nanoparticles and individual biomolecules [1]. Here we show that this nanothermodynamics may also apply to large systems if each particle interacts with a finite number of neighboring particles, so that neighboring particles form a local region that fluctuates independent of the infinite bath. Indeed, a variety of measurement techniques have shown that independently relaxing regions govern the primary response of many materials [2, 3], including ferromagnets [4-6], at least in the paramagnetic phase above the critical temperature ($T_C$) on which we focus here. A possible mechanism comes from the constraint necessary to conserve local momentum while changing the state of a particle. Similar constraints are the basis of the theory of fluctuations for a small part of a large system [7] and Onsager's irreversible thermodynamics [8, 9]. We extend these ideas to the fluctuations of individual particles in thermal equilibrium. Previous simulations using the Ising model have shown that a specific fluctuation constraint lowers the energy, and makes the entropy additive and extensive [10], consistent with a primary postulate of nanothermodynamics, and with measurements [11-13] and analysis [14,15] of non-resonant spectral hole burning in many materials. Here we show that this constraint greatly improves the agreement between the Ising model and measured susceptibility of ferromagnetic materials and critical fluids.

In 1872 Boltzmann introduced his fundamental expression for statistical entropy, $S=k_B \ln \Omega$. Here $k_B$ is Boltzmann's constant, and $\Omega$ is the number of available microstates, which for magnetic systems is a function of energy $E$ and magnetic alignment $M$. Inverting this expression yields the probability $p \propto \exp[\Delta S / k_B]$ for a change in total entropy, $\Delta S$. Although changing the state of a single particle (subscript 1) may change its energy $\Delta E_1$ and/or alignment $\Delta M_1$, because the change is between specific microstates, the entropy change of the particle is zero $\Delta S_1 = 0$. Thus, in the standard canonical ensemble, all of the change in entropy comes from the infinite bath $\Delta S^* = \frac{1}{T}\Delta E^* - \frac{H}{T}\Delta M^*$. Here the temperature ($T$) and external field ($H$) are defined by first-order derivatives of the entropy, $\frac{\partial S^*}{\partial E^*} = \frac{1}{T}$ and $\frac{\partial S^*}{\partial M^*} = -\frac{H}{T}$. We now examine the conservation laws that govern the change in total entropy, and give the probability $p \propto \exp[\Delta S / k_B]$ for this change.

Conservation of energy in the canonical ensemble requires thermal contact between the single particle and the infinite bath, Fig. 1a, so that $\Delta E^* = -\Delta E_1$. Because $\Delta S_1 = 0$, the change in total entropy is $\Delta S = \Delta S^* = -\Delta E_1/T$. Thus, the usual Boltzmann factor $p \propto \exp[-\frac{\Delta E_1}{k_B T}]$ comes from the linear change in entropy of the infinite bath due to heat exchange with the single particle. Nonlinear terms are negligible because the infinite bath absorbs changes in energy without significantly altering the bath.

Conservation of momentum often requires direct interactions between particles, Fig. 1b. Specifically, inverting the alignment (angular momentum) of a magnetic spin requires a magnetic field, which for $H=0$ may come from the exchange interaction between neighbors. Similarly, changing the linear momentum of a particle requires a net force, which comes primarily from nearby neighbors. Furthermore, changing the state of the single particle changes its neighbors differently than it changes more distant particles. For example, if the potential energy of the single particle decreases, the potential energy of the interacting neighbors will also decrease, opposite to the energy of the infinite bath which must increase to conserve energy. Thus, the environment of an interacting particle often includes a finite-sized local region in addition to the infinite bath, as confirmed by measurements of independently relaxing regions. The statistical independence of the local regions ensures that their equilibrium entropies are additive in the total entropy, so that the probability of changing a particle's state is again $p \propto \exp[\Delta S / k_B]$, but now with $\Delta S = \Delta S_1 + \Delta S^* + \Delta S_L$. The first two terms are the same as for the canonical ensemble, $\Delta S_1 = 0$ and $\Delta S^* = \Delta E^*/T$. The additional term ($\Delta S_L$) is the offset from equilibrium entropy of the local region due to its fluctuations in energy ($\Delta E_L$) and alignment ($\Delta M_L$). In thermal equilibrium, the first-order derivatives of $S_L$ are equal to those of $S^*$: $\partial S_L/\partial E_L = 1/T$ and $\partial S_L/\partial M_L = 0$ when $H=0$. Two of the second-order derivatives can be written as $\partial^2 S_L/\partial E_L^2 = -1/(T^2 C_L)$ and $\partial^2 S_L/\partial M_L^2 = -1/(N_L T \chi_L)$, with $C_L = (<E_L^2> - <E_L>^2)/k_B T^2$ the local heat capacity and $\chi_L = (<M_L^2> - <M_L>^2)/N_L k_B T$ the local magnetic susceptibility. Here $N_L$ is the number of particles in the region, which is chosen to minimize the average energy per particle, as described below. Thus, the change in entropy of the region to second order is:

$$\Delta S_L = \frac{\Delta E_L}{T} - \frac{(\Delta E_L)^2}{2T^2 C_L} + \frac{\partial^2 S_L}{\partial E_L \partial M_L} \Delta E_L \Delta M_L - \frac{(\Delta M_L)^2}{2N_L T \chi_L}. \qquad \text{Eq. (1)}$$

First focus on the linear terms in the change of total entropy, which includes only the first term on the right side of Eq. (1), $\Delta S \approx 0 + \Delta E^*/T + \Delta E_L/T$. Again using conservation of energy $\Delta E^* = -\Delta E_1 - \Delta E_L$, the entropy change to first order is $\Delta S \approx -\Delta E_1/T$, giving the same

Boltzmann factor as before: $p \propto \exp[-\frac{\Delta E_1}{k_B T}]$. However, because the local region is finite, nonlinear contributions to the entropy change are significant. We focus on the final term in Eq. (1), which comes from fluctuations in alignment. We neglect higher-order terms in energy partly for simplicity, but also because the linear term keeps $E_L$ near its equilibrium value. Indeed, from simulations of the Ising model adjusted to the optimal local region size ($N_L = 27$) and constraint (as described below), we find that the second order term in $\Delta M_L$ is much larger than that for $\Delta E_L$ ($|\partial^2 S_L/\partial M_L^2| > 4.5 |\partial^2 S_L/\partial E_L^2|$), thus justifying our neglect of $\partial^2 S_L/\partial E_L^2$. Moreover, when $H=0$, $\partial^2 S_L/\partial M_L^2$ gives the lowest-order nonzero term for changes in alignment, so that this term must be included if the local region is to serve as a thermodynamic bath for alignment change.

An approximate expression for the entropy of alignment of Ising particles comes from the binomial coefficient $S_L = k_B \ln\{N_L!/[\frac{1}{2}(N_L + \Delta M_L)]![\frac{1}{2}(N_L - \Delta M_L)]!\}$. Using Stirling's formula for the factorials gives $\frac{\partial^2 S_L}{\partial M_L^2} = -k_B/N_L$, which when put into Eq. (1) yields the modified Boltzmann factor:

$$p \propto \exp[-\frac{\Delta E_1}{k_B T} - g\frac{(\Delta M_L)^2}{2N_L}(1 - \delta_{\Delta E_1 0})]. \qquad \text{Eq. 2}$$

Here $g$ is a constraint parameter, while the Kronecker delta ensures that if there is no net interaction between particles ($\Delta E_1 = 0$), then the factor $(1 - \delta_{\Delta E_1 0}) = 0$ removes the fluctuation constraint. Simulations without this $\delta_{\Delta E_1 0}$ are overly constrained, giving a critical temperature that is several times larger than the usual $T_C$, supporting the argument that the fluctuation constraint applies only when particles interact directly. If $\Delta E_1 \neq 0$, the constraint parameter controls the strength of the constraint: for example, $g=0$ gives the usual Boltzmann factor, $g=1$ gives the normal Gaussian constraint expected from fluctuation theory, and $g\rightarrow\infty$ prevents the particle from changing its state unless there is no net alignment in the region or no net interaction with neighboring particles. The constraint is a type of entropic force, similar to temperature, but comes from a second-order term that is significant if a particle interacts directly with a finite number of nearby particles. Specifically, when a region of interacting particles fluctuates into a partially-aligned state $\Delta M_L \neq 0$, because there is less entropy available, the probability of inverting a particle in the region is reduced below that of the Boltzmann factor alone. Monte-Carlo simulations of the Ising model show that for $g\sim 1$ the average energy is minimized, while fluctuations in energy are maximized, making the entropy homogeneous, additive, and extensive [10]. This $g\sim 1$ also gives an energy dependence that is inversely proportional to $N_L$, consistent with a mean-field cluster model and Landau theory for the non-exponential and non-Arrhenius response measured in many materials [16, 17]. The model and theory have been used to describe non-classical critical scaling in ferromagnets [18, 19]. Here we show that the Ising model with this fluctuation constraint provides a microscopic picture for the measured corrections to scaling in ferromagnetic materials and simple fluids near their critical points.

We investigate the influence of the fluctuation constraint in Eq. (2) using Monte-Carlo simulations of the 3-dimensional Ising model. We start with a 24x24x24 simple-cubic lattice with periodic boundary conditions. The particle at site $i$ may be "up" ($\sigma_i=+1$) or "down" ($\sigma_i=-1$), which simulates the two alignments of a uniaxial magnet. Other systems that can be simulated using the Ising model include binary alloys, critical fluids, and glass-forming liquids where $\sigma_i$ may represent the relative fraction of two competing structures [17]. The net energy per particle is $E/N = -\frac{1}{2} J \sum_{i=1}^{N} \sigma_i H_i / N$. Here the sum is over all $N=(24)^3$ particles in the lattice, the factor of ½ removes double counting, and the exchange constant is $J>0$ for ferromagnetic interactions. The local field at each site due to direct interactions is $H_i = \Sigma_{<ij>} \sigma_j$, where the sum is over all six nearest neighbors. The thermal equilibrium energy per particle is $<E>/N = -\frac{1}{2} J <\sum_{i=1}^{N} \sigma_i H_i> / N$, and the susceptibility is $<\chi> = [<(\sum_{i=1}^{N} \sigma_i)^2> - (<\sum_{i=1}^{N} \sigma_i>)^2] / Nk_BT$, where the averages are found by simulating over $10^6$ Monte-Carlo sweeps. Results are evaluated as a function of reduced temperature, $\tau = (T-T_C)/T_C$.

We investigate the fluctuation constraint by subdividing the Ising lattice into $N/N_L$ equal-sized cubic-shaped regions, with $N_L=8$, 27, or 64 particles. Each particle interacts with all six of its nearest neighbors, including neighbors that are outside its region. Thus, the only influence of subdividing the lattice comes from the nonlinear term in Eq. (2), which depends on the instantaneous alignment of the region $M_L = \sum_{i=1}^{N_L} \sigma_i$, with strength governed by $g$. If $g=0$, Eq. (2) reduces to the usual Boltzmann factor, yielding standard Ising behavior where the accepted critical temperature is $k_B T_C/J=4.51152$. If $g>0$, the constraint lowers the energy, increases the fluctuations, and raises the critical temperature. The critical temperatures used here, chosen to be consistent with the temperature at which the heat capacity is a maximum and to match data as $\log(\tau) \to -2$, are: $k_B T_C/J=4.509$, 6.102, 6.670, and 6.674 for $g = 0.0$, 0.5, 1.0, and 2.0, respectively. Note that 4.509 for $g=0$ is within simulation uncertainty (0.1 %) of the standard value. Also note that simulations at $T<T_C$ using Eq. (2) with $g>0$ exhibit unrealistic behavior because a different constraint applies in the ferromagnetic phase where the equilibrium alignment is nonzero.

Figure 2 shows some results of the fluctuation constraint on the Ising model. The upper right inset is a plot of the average energy per particle as a function of $g$ at four temperatures. The minimum in $<E>/N$ occurs at $g\sim2$, somewhat above $g=1$ expected from the binomial coefficient. Although some variation could come from higher-order terms that are not included in Eq. (2), another consideration is that the true thermal equilibrium comes from minimizing the free energy, not just $<E>/N$. Moreover, $g\sim1$ maximizes the local fluctuations in energy, which makes the entropy homogeneous [10]. For $g=1$, the lower left inset shows $<E>/N$ as a function of region size. The region size with lowest $<E>/N$ changes from $N_L=8$ to $N_L=27$ at $k_BT/J=8.24$, but remains at $N_L=27$ as $T\rightarrow T_C$. Here, for simplicity, we use a fixed value of $N_L=27$ for comparison with measurements. More realistic behavior involves a distribution of region sizes with an average $<N_L>$ that varies continuously with temperature, as is found from the generalized ensemble in the mean-field cluster model and Landau theory [16-19]. Future plans call for developing simulations that adjust the size of the regions in a self-consistent manner.

The main part of Fig. 2a is a critical-scaling (log-log) plot of susceptibility versus reduced temperature. The symbols show the measured susceptibility of Gd ($\chi_{Gd}$) [20 -23]. Because only a few ferromagnetic materials have been found that agree with standard models as $T\rightarrow T_C$ [24, 25], here we focus on corrections to scaling at $\log(\tau) > -2$ where long-range interactions, local disorder, and uncertainty in the determination of $T_C$ have less impact. The dashed curve in Fig. 2a shows that $<\chi_0>$ from the standard Ising model with $g=0$ has a slope that is significantly steeper than the slope of $\chi_{Gd}$ over most of this temperature range. Whereas the solid curve in Fig. 2a shows that $<\chi_1>$ from the Ising model with $g=1$ and $N_L=27$ gives better agreement. Figure 2b is a residual plot. The dashed curve shows a significant difference between log($<\chi_0>$) and log($\chi_{Gd}$), whereas the symbols show no systematic deviation between log($<\chi_1>$) and log($\chi_{Gd}$) over more than two orders of magnitude in $\chi$ and $\tau$. We emphasize that $<\chi_1>$ is not a fit to the data, but comes from the Ising model with optimal constraint and no adjustable parameters.

Figure 3 shows the effective scaling exponent $\gamma_{eff} = -d\log(\chi)/d\log(\tau)$ as a function of reduced temperature. The symbols come from the measured susceptibility of two magnetic materials (Gd and Ni [26, 27]) and one critical fluid ($CO_2$ [28]), while the lines come from Monte-Carlo simulations of the Ising model. Noise in the simulations and data was reduced by smoothing and taking the derivative using a Savitzky-Golay filter. The data show three distinct features that are not found in the standard Ising model with $g=0$ (dashed curve). The data have an interval of relatively constant $\gamma_{eff}$ above $\log(\tau)=-1$, whereas the dashed curve changes continuously. Over this interval the data have $\gamma_{eff} \approx 1.1$, whereas the dashed curve has $\gamma_{eff} > 1.2$. Below $\log(\tau)=-1$ the data show a sharp rise in $\gamma_{eff}$, which is not in the standard Ising model with $g=0$. Some substances show a smaller rise, which may be simulated by adjusting the strength of the constraint between $g=0.5$ (dash-dot curve) and $g=2.0$ (dash-dot-dot curve). Adjusting the value of $g$ may indicate that additional nonlinear terms should be included in Eq. (2).

Several other models have been used to try and explain the measured corrections to scaling. A uniaxial dipolar model has been proposed for Gd [29], but this model predicts a sharp decrease from $\gamma_{eff}=1.24$ at $\log(\tau)=0$ to $\gamma_{eff}<1.1$ at $\log(\tau)=-2$, opposite to the sharp increase shown by the data in Fig. 3. Although a crossover to uniaxial-dipolar behavior could explain the asymptotic value of $\gamma_{eff} \rightarrow 1.0$ at $\log(\tau) < -2$ [20], the mean-field cluster model [18], Landau theory [19], and simulations of the Ising model with $g=1$ (described here but not shown) all have similar shifts towards mean-field scaling as the transition is approached. A shift towards mean-field behavior has also been used to explain the empirical scaling exponent of Ni [26], but again because most measurements on ferromagnetic materials do not agree with standard models for asymptotic critical behavior [24, 25], here we focus on corrections to scaling at log($\tau$)>–2. Standard models predict corrections to scaling that depend on sample details, such as spin value and interaction range [30, 31], but all measurements shown in Fig. 3 have similar values for $\gamma_{eff}$ near log($\tau$) = –1. Furthermore, most standard models in the canonical ensemble do not exhibit the sharp rise in $\gamma_{eff}$ found for the ferromagnetic materials.

The inset of Fig. 3 shows the normalized effective Weiss constant ($\theta_{eff}/T_C$) as a function of $g$. The standard Weiss constant $\theta$ in the Curie-Weiss law $\chi \propto 1/(T-\theta)$ is given by the temperature at which $1/\chi = 0$ when extrapolated from a linear fit of $1/\chi$ at high temperatures. But Fig. 3 shows that the data never fully obey the Curie-Weiss law. Thus, to be precise, we define an effective Weiss constant ($\theta_{eff}$) by the temperature at which $1/\chi = 0$ when extrapolated from a linear fit of $1/\chi$ over the temperature range $-1 \leq$ log($\tau$) $\leq -0.5$. The ratio $\theta_{eff}/T_C$ is well-defined experimentally, and is relatively insensitive to any offset in temperature. The open symbols in the inset of Fig. 3 are from simulations of the Ising model, showing that $\theta_{eff}/T_C$ has a maximum as $g \rightarrow 0$ and a minimum at $g \approx 1$. The solid symbols are from the measured susceptibility of three ferromagnetic materials (Gd, Ni, and EuO [32,33]) and two critical fluids ($CO_2$ and Xe [28]), showing that $\theta_{eff}/T_C$ deviates significantly from the standard Ising model with $g=0$.

It has been argued that a more fundamentally consistent effective scaling exponent comes from the nonlinear reduced temperature $\tau' = \tau T_C/T = (T-T_C)/T$ [34-36].

Figure 4 shows the nonlinear effective scaling exponent $\gamma_{eff}$'=–dlog($\chi T$)/dlog($\tau$') as a function of log($\tau$'). The symbols are from the five substances shown in Fig. 3. The dashed curve is from the Ising model with $g$=0, showing the failure of this model to match measured data using standard Boltzmann statistics. The solid curve is from the Ising model with $g$=1, showing excellent agreement with several features in the Gd data, and good agreement with the other substances at temperatures not too close to the transition. For example, note the quantitative agreement between the critical fluid data and the Ising model with $g$=1 over the range –1.4 < log($\tau$') < –0.7. Thus these fluids, which are known to belong to the Ising universality class, show clear deviations from the standard Ising model above $T_C$. Such deviations from standard Ising behavior have been attributed to a structural length scale [37] that coexists with the usual correlation length, $\xi$. Indeed, evidence for a second length scale near the critical point of many materials comes from high-resolution x-ray and neutron scattering [38, 39]. In our simulations the second length scale comes from the region size via $(N_L)^{1/3}$. Some consequences of this length scale can be deduced from features in the temperature-dependent susceptibility near the crossover where the optimal region size ($N_L$=27 lattice sites) equals the correlation volume $4\pi\xi^3/3$. Note that we neglect the temperature dependence of $N_L$, which is weak compared to the divergence in $\xi$, as can be inferred from the lower-left inset of Fig. 2a. Using the atomic volume and temperature dependence of $\xi$ [25, 40], we find $4\pi\xi^3/3$=27 lattice sites at log($\tau$') = –0.8 for Ni, and at log($\tau$') ~ –1 for other ferromagnetic materials. Thus, log($\tau$')~–1 marks the temperature where long-range (inter-regional) correlations change sharply. Qualitatively, sharp changes in the slope of the susceptibility indicate that some higher-order transition occurs as a precursor to the usual critical point.

Quantitatively, the solid curves in Figs. 2-4 show that the Ising model with $g=1$ accurately simulates several details in the measured corrections to scaling.

In summary, we have shown that the standard Ising model fails to describe the measured susceptibility of several substances, whereas adding a nonlinear constraint from fluctuation theory greatly improves agreement. We present arguments that the nonlinear constraint comes from finite-sized local regions that form inside the infinite bath. A possible mechanism involves the local interactions necessary to change the state of a particle while conserving momentum. The nonlinear constraint lowers the net free energy, and yields a distribution of energies compatible with the dynamical heterogeneity found in a wide range of materials. Moreover, the optimal constraint makes the local entropy homogeneous, additive, and extensive, consistent with a fundamental postulate of nanothermodynamics. This optimal constraint significantly alters the magnitude and temperature dependence of the thermal-equilibrium susceptibility, and increases the critical temperature by nearly 50 %, indicating that calculations of interparticle interactions and corrections to scaling must be modified to compare with data at finite temperatures. Because the constraint comes from standard fluctuation theory for a small part of a large system, the true thermal equilibrium of many systems of interacting particles may involve a significant nonlinear component, beyond the Boltzmann factor.

We thank S. Srinath and S. N. Kaul for providing their original data; and E. Bauer, N. Bernhoeft, I. A. Campbell, R. Richert, and L. Tu for helpful discussions. We thank the Army Research Office for financial support; and The High Performance Computing Initiative at ASU for technical assistance.

Figure Captions

Fig. 1. Cartoon sketch of relevant thermodynamic systems. **a**, In the usual canonical ensemble, conservation of energy is maintained while changing the state of a single particle (+) by thermal contact to an infinite heat bath having entropy $S^*$ at energy $E^*$. **b**, Conservation of momentum may require direct interactions between the particle and a finite number of neighboring particles, forming a local region that is distinct from the infinite bath. The local region provides a local heat bath having entropy $S_L$ at energy $E_L$. Furthermore, due to finite-size effects, $S_L$ depends nonlinearly on the local alignment $M_L$, so that the local region may also provide an alignment bath for the particle.

Fig. 2. **a**, Log-log (scaling) plot of magnetic susceptibility versus reduced temperature, $\tau=(T-T_C)/T_C$. Open symbols (□) are from measurements of Gd ($\chi_{Gd}$) [20-23], multiplied by a factor to give an amplitude similar to the simulations. The dashed line comes from simulations of the Ising model using the Boltzmann factor alone $g=0$ ($<\chi_0>$), while the solid line comes from the Boltzmann factor with fluctuation constraint $g=1$ ($<\chi_1>$). Typical uncertainties in $<\chi_0>$ and $<\chi_1>$ are 2 %. The upper right inset shows the average energy per particle ($<E>/N$) as a function of the constraint parameter ($g$), with the number of particles in the local region $N_L=27$ at four temperatures ($k_BT/J=10$-■, 8.5-●, 7.5-▲, and 7.0-▼). The lower left inset shows $<E>/N$ as a function of $N_L$ with $g=1$ at the same four temperatures. Typical uncertainties in $<E>/N$ are 0.1 %. Lines connecting the symbols are a guide for the eye. **b**, Residual plot from **a**, showing $\log(<\chi_1>)-\log(\chi_{Gd})$ (symbols) and $\log(<\chi_0>)-\log(\chi_{Gd})$ (dashed curve). Linear interpolation was used to obtain the $\chi$ at equally-spaced values of log($\tau$). Although the data have been multiplied by a factor, simulations of $\langle\chi_1\rangle$ use the optimal values of $g=1$ and $N_L=27$ with no adjustable parameters.

Fig. 3. Effective scaling exponent $\gamma_{eff} = -d\log(\chi)/d\log(\tau)$ (negative slope when plotted as in Fig. 2a) as a function of reduced temperature. Symbols are from measured susceptibilities of two magnetic materials (Gd-■ from Fig. 2 and Ni-● [26,27]), and one critical fluid ($CO_2$-▲[28]). Lines are from simulations of the Ising model using: the standard Boltzmann factor $g=0$ (dashed); and fluctuation constraint with constraint parameters $g=0.5$ (dash-dot), $g=1$ (solid), and $g=2$ (dash-dot-dot). Inset shows the normalized effective Weiss constant as a function of constraint parameter for the Ising model-□, three magnetic materials (Gd-■, Ni-●, and EuO-♦ [32, 33]), and two critical fluids ($CO_2$-▲ and Xe-▼ [28]). Values of $g$ for the data are estimated from the qualitative behavior of the effective scaling exponent.

Fig. 4. Nonlinear effective scaling exponent $\gamma_{eff}' = -d\log(\chi T)/d\log(\tau')$ as a function of nonlinear reduced temperature, $\tau' = (T-T_C)/T$. Symbols are from measured susceptibilities of three magnetic materials (Gd-■, Ni-●, and EuO-♦), and two critical fluid ($CO_2$-▲ and Xe-▼). Lines are from simulations of the Ising model using: the standard Boltzmann factor $g=0$ (dashed); and fluctuation constraint with optimal constraint parameter $g=1$ (solid).

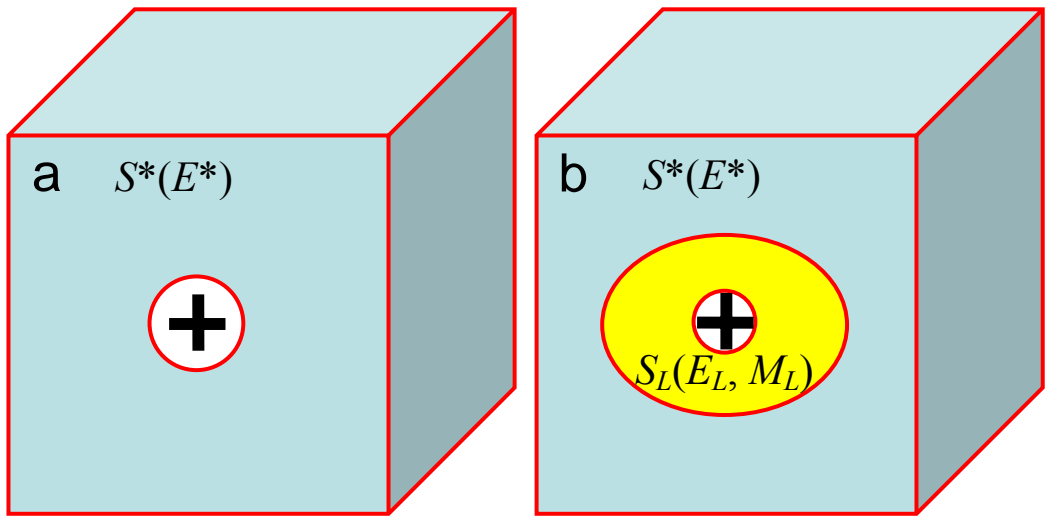

Fig. 1

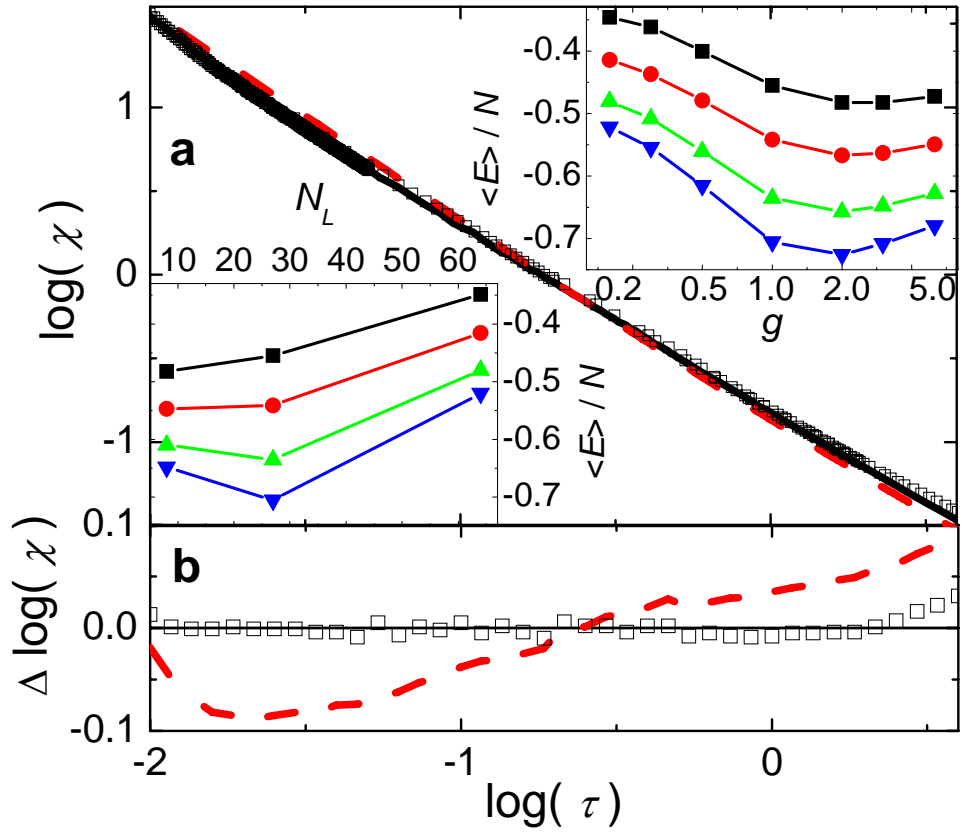

Fig. 2
19

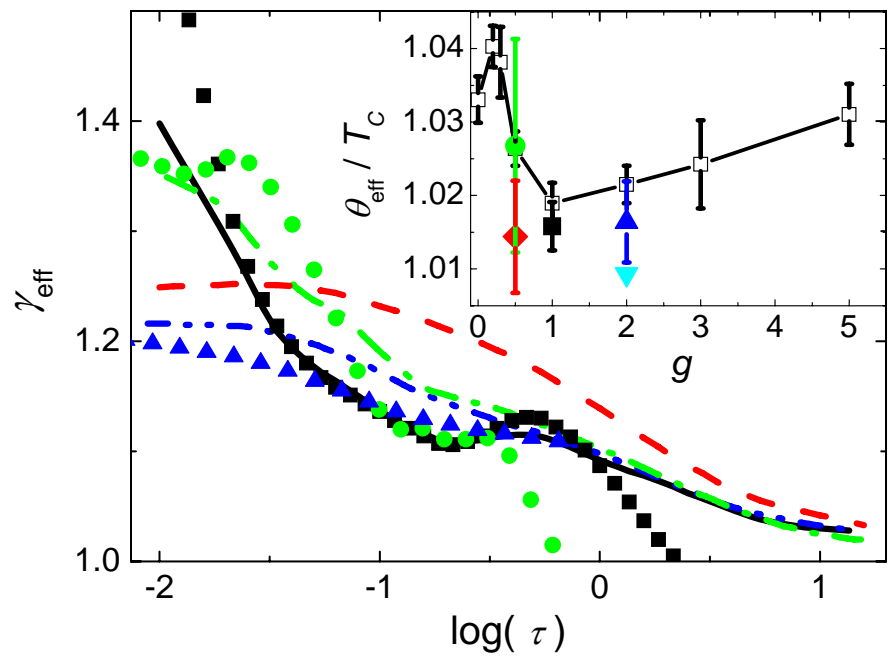

Fig. 3

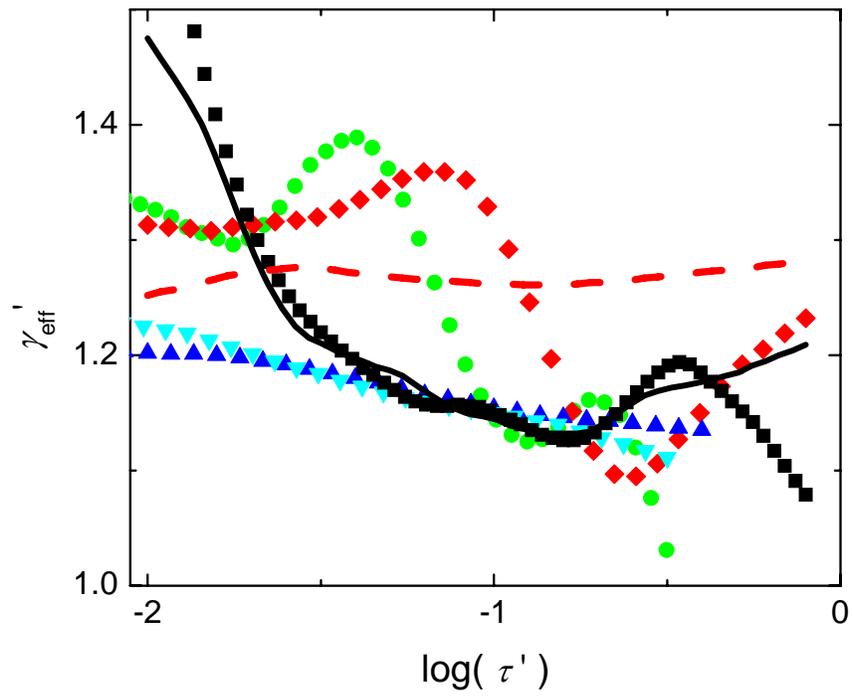

Fig. 4